# Space Charge in Circular Machines

*M. Ferrario, M. Migliorati, and L. Palumbo*
INFN-LNF and University of Rome "La Sapienza"

**Abstract**
Space charge forces, which arise directly from the beam's charge distribution and include the influence of image charges and currents induced by interactions with a perfectly conducting, smooth pipe, are very important in high-intensity, low-energy accelerators. These forces play a key role under various beam dynamics regimes, leading to effects such as energy spread, modifications of betatron tunes, and potential instabilities. This lecture will explore the fundamental characteristics of space charge effects in circular accelerators.

## 1. Introduction

Charged particles in accelerators are steered, confined, and accelerated by external electromagnetic (e.m.) fields. Specifically, acceleration is driven by the electric field in RF cavities, while magnetic fields provide guidance and focusing. Bending magnets direct the particles along the reference trajectory (orbit), while solenoids and quadrupoles ensure transverse confinement. Additionally, sextupoles are used to correct chromaticity.

To derive the equation of motion under the effects of these e.m. fields, we can start from Newton second law with the Lorentz force through the equation:

$$\frac{d(m_0 \gamma \vec{v})}{dt} = \vec{F}^{ext} = e(\vec{E} + \vec{v} \times \vec{B}), \tag{1}$$

where $m_0$ is the electron rest mass, $\gamma$ is the relativistic factor, $e$ is the electron charge and $\vec{v}$ is the particle velocity. In principle, the use of the above equation allows us to determine the trajectory of a charge in any e.m. field.

The external force $\vec{F}^{ext}$ is independent of the beam current. It can be generally controlled through the magnets' currents and the RF power. In an accelerator, however, there is another source of e.m. fields that becomes particularly important at high currents. It is the beam itself, which, circulating inside the pipe, produces additional e.m. fields called "self-fields". These fields depend on the beam intensity and distribution, and on the surrounding environment, and they can perturb the external guiding fields.

In general, they are responsible for several unwanted phenomena related to beam dynamics, such as beam shape distortion, energy loss and spread, shifts of the synchronous phase and frequency, shifts of the betatron frequencies, and, under some conditions, instabilities. It is customary to divide the study of self-fields into space charge fields and wakefields. The space charge forces are those generated directly by the Coulomb interaction in a multi-particle system as a beam, or by the image charges and currents originated by the interaction of the beam with a perfectly conducting smooth pipe [1]. The wakefields are instead due to the finite conductivity of the walls and to all geometric variations of the beam pipe (such as resonant devices and transitions of the beam pipe). A reference paper on the wakefields can be found in [2].

## 2. Collisional and space charge regimes

This lecture only discusses the effects of the space charge forces. In particular, when dealing with the Coulomb interaction in a multi-particle system, we can distinguish two regimes [3]:

- *collisional regime*, in which a charge motion is dominated by binary collisions with other single particles.

- *collective regime or space charge regime*, in which a charge motion is dominated by the self-field produced by the whole particle distribution which varies appreciably only over large distances compared to the average separation of the particles.

The collisional part of the total interaction force can cause random displacements of the particle's trajectory and statistical fluctuations in the particle distribution, leading for example to intra-beam scattering effects in high-energy storage rings [4] (see also Touschek effect [5]). On the other hand, space charge forces lead to the collective behavior of the beam driving for example envelope oscillations, emittance and energy spread growth [6].

A way to quantify the importance of collisional or collective effects in a beam is the Debye length [3] $\lambda_D = \sqrt{\frac{\varepsilon_o \gamma^2 k_B T}{e^2 n}}$, where $\varepsilon_0$ is the dielectric constant of vacuum, $n$ is the particle density and the transverse beam temperature $T$ is defined as $k_B T = \gamma m_o \langle v_\perp^2 \rangle$, $k_B$ being the Boltzmann constant. If a test charge is placed inside the beam, the excess electric potential $\Phi_D$ set up by this charge is effectively screened off in a distance $\lambda_D$ by charge redistribution in the beam as: $\Phi_D(\vec{r}) = \frac{C}{r} e^{-r/\lambda_D}$. This effect is well known from plasma physics as *Debye shielding* [7]. A charged particle beam in a particle accelerator can be viewed in fact as a non-neutral plasma [8] in which the smooth focusing channel replaces the restoring force produced by ions in a neutral plasma, see Fig. 1. Like in a neutral plasma, collective behavior of the beam can be observed on length-scales much larger than the Debye length. It follows that if the Debye length is much smaller compared to the beam radius, collective effects generated by smooth forces due to the self fields of the beam will play a dominant role in driving the beam dynamics with respect to binary collisions.

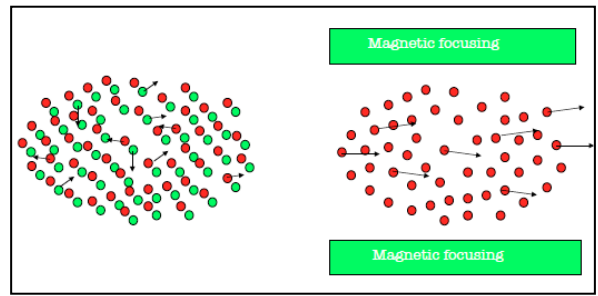

**Fig. 1:** The restoring force produced by the ions (green dots) in a neutral plasma can be replaced by a smooth focusing channel for charged particle beam (non-neutral plasma) in a particle accelerator.

Smooth functions for the charge and field distributions can be used, as will be done in paragraph 3, as long as the Debye length remains large compared to the interparticle distance $d = n^{-1/3}$, that is as long as the number $N_D$ of particles within a Debye sphere of radius $\lambda_D$ remains large ($N_D \gg 1$). A typical particle is actually scattered by all of the other particles within its Debye sphere, but the large number of random interactions very rarely causes any sudden change in its motion (weakly coupled plasma) and mainly contributes to driving the beam toward a thermal equilibrium [3].

Notice that the Debye length increases with particle energy γ so that at sufficiently high energy a transition from space charge to collisional regime may occur.

## 3. Self fields and equations of motion

### 3.1 The betatron motion

Before dealing with the self-induced forces produced by the space charge and their effect on the beam dynamics in a circular accelerator, we briefly review the transverse equations of motion [9]. In order to simplify our study, let us consider a perfectly circular accelerator with radius $\rho_x$ and obtain the transverse single-particle equation of motion in the linear regime.

If we include, in the particle equation of motion given by Eq. (1), the self-induced forces, we have

$$\frac{d(m_0\gamma\vec{v})}{dt} = \vec{F}^{ext}(\vec{r}) + \vec{F}^{self}(\vec{r}). \tag{2}$$

By considering a constant energy γ it becomes:

$$\frac{d\vec{v}}{dt} = \frac{\vec{F}^{ext}(\vec{r}) + \vec{F}^{self}(\vec{r})}{m_0\gamma} . \tag{3}$$

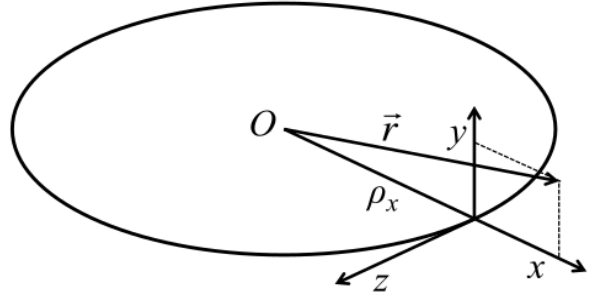

**Fig. 2:** Coordinate system for a charge in a circular accelerator.

According to the coordinate system of Fig. 2, indicating with $\vec{r}$ the charge position, and with $x$ and $y$ the transverse displacements with respect to the reference trajectory, we write:

$$\vec{r} = (\rho_x + x)\hat{e}_x + y\hat{e}_y. \tag{4}$$

Since the unit vector $\hat{e}_x$ rotates with angular frequency $\omega_0$ (clockwise in the figure), its time derivative is $\omega_0 \hat{e}_z$, so that the velocity is

$$\vec{v} = \frac{d\vec{r}}{dt} = \dot{x}\hat{e}_x + \dot{y}\hat{e}_y + \omega_0(\rho_x + x)\hat{e}_z \tag{5}$$

and the acceleration

$$\vec{a} = \left[\ddot{x} - \omega_0^2(\rho_x + x)\right]\hat{e}_x + \ddot{y}\hat{e}_y + \left[\dot{\omega}_0(\rho_x + x) + 2\omega_0\dot{x}\right]\hat{e}_z . \tag{6}$$

In the above equations the dots mean derivative with respect to time.

If we consider the motion along x, we have:

$$\ddot{x} - \omega_0^2 (\rho_x + x) = \frac{1}{m_0 \gamma} \left( F_x^{ext} + F_x^{self} \right). \tag{7}$$

Instead of using the time *t* as variable, we consider the azimuthal position $s=v_z t$, so that the acceleration along *x* becomes

$$\ddot{x} = \frac{d^2 x}{dt^2} = v_z^2 \frac{d^2 x}{ds^2} = v_z^2 x'' = \omega_o^2 (\rho_x + x)^2 x'' \tag{8}$$

for which we have also used Eq. (5). With the above equation, the differential equation of motion (7) can be written as

$$x'' - \frac{1}{\rho_x + x} = \frac{1}{m_0 v_z^2 \gamma} \left( F_x^{ext} + F_x^{self} \right). \tag{9}$$

We assume now small transverse displacements *x* with respect to the machine radius $\rho_x$, so that we can linearly expand the above equation into

$$x'' - \frac{1}{\rho_x} + \frac{1}{\rho_x^2} x = \frac{1}{m_0 v_z^2 \gamma} \left( F_x^{ext} + F_x^{self} \right). \tag{10}$$

The external forces are produced by the magnetic guiding fields. We suppose to have only a linear lattice composed of dipoles and quadrupoles. This means that we can expand the external guiding fields in a Taylor series up to the quadrupole component

$$-F_x^{ext} = q v_z B_y = q v_z B_{y0} + q v_z \left( \frac{\partial B_y}{\partial x} \right)_0 x + \cdots \tag{11}$$

and recognize that the dipolar magnetic field $B_{y0}$ is responsible for the circular motion along the reference trajectory of radius $\rho_x$ according to the equation

$$q v_z B_{y0} = \frac{m_0 \gamma v_z^2}{\rho_x}. \tag{12}$$

We finally obtain

$$x'' + \left[ \frac{1}{\rho_x^2} + \frac{q}{m_0 v_z \gamma} \left( \frac{\partial B_y}{\partial x} \right) \right] x = \frac{1}{m_0 v_z^2 \gamma} F_x^{self}, \tag{13}$$

which can also be written as

$$x'' + \left[ \frac{1}{\rho_x^2} - k \right] x = \frac{1}{m_0 v_z^2 \gamma} F_x^{self}, \tag{14}$$

where we have introduced the normalized gradient

$$k = \frac{g}{p/q} = -\frac{q}{m_0 v_z \gamma} \left( \frac{\partial B_y}{\partial x} \right) \tag{15}$$

with $g$ the quadrupole gradient in [T/m] and $p$ the charge momentum.

We observe that both the curvature radius $\rho_x$ and the normalized gradient depend on the azimuthal position 's' along the accelerator. By using the focusing constant $K_x(s)$ we then can write

$$x''(s) + K_x(s)x(s) = \frac{1}{m_0 v_z^2 \gamma} F_x^{self}(x,s). \tag{16}$$

In the absence of self-fields, Eq. (16) is known as Hill's equation, and its solution gives the well-known betatron oscillations of the kind:

$$x(s) = a_x \sqrt{\beta_x(s)} \cos[\mu_x(s) - \varphi_x], \tag{17}$$

where $a_x$ and $\varphi_x$ depend on the initial conditions, and

$$\begin{aligned} &\frac{1}{2}\beta_x \beta_x'' - \frac{1}{4}\beta_x'^2 + K_x(s)\beta_x^2 = 1 \\ &\mu_x'(s) = 1/\beta_x(s) \\ &Q_x = \frac{\omega_x}{\omega_o} = \frac{1}{2\pi}\int_0^L \frac{ds'}{\beta_x(s')} \end{aligned}, \tag{18}$$

where we have defined the betatron tune $Q_x$ as the number of transverse oscillations per revolution.

In the analysis of the motion in the presence of the self-induced fields, it is convenient to adopt a simplified model approximating the betatron motion with harmonic oscillations around the reference trajectory. This is equivalent to having a constant focusing term $K_x$ along the machine. Although this case has never been fulfilled in a real accelerator, however, it provides a reliable model for the description of the beam instabilities. Under this approximation Eq. (16) becomes

$$x''(s) + K_x x(s) = \frac{1}{m_0 v_z^2 \gamma} F_x^{self}(x,s), \tag{19}$$

which is a linear differential equation. The homogeneous solution is given by

$$x(s) = A_x \cos\left[\sqrt{K_x} s - \varphi_x\right] \tag{20}$$

were, with the notations of Eq. (18), we have

$$\begin{aligned} &a_x \sqrt{\beta_x} = A_x \\ &\beta_x = \frac{1}{\mu_x'} = \frac{1}{\sqrt{K_x}} \\ &\mu_x(s) = \sqrt{K_x} s \\ &Q_x = \frac{1}{2\pi}\int_0^L \frac{ds'}{\beta_x(s')} = \frac{L}{2\pi\beta_x} = \rho_x \sqrt{K_x} \Rightarrow K_x = \left(\frac{Q_x}{\rho_x}\right)^2 \end{aligned}. \tag{21}$$

The differential equation of motion (19) then becomes

$$x''(s) + \left(\frac{Q_x}{\rho_x}\right)^2 x(s) = \frac{1}{m_0 v_z^2 \gamma} F_x^{self}(x,s) \ . \tag{22}$$

An analogous equation of motion can also be written for the vertical plane:

$$y''(s) + \left(\frac{Q_y}{\rho_x}\right)^2 y(s) = \frac{1}{m_0 v_z^2 \gamma} F_y^{self}(y,s) \ . \tag{23}$$

Equations (22) and (23) represent our starting point for studying the effects of the self-induced fields on the betatron oscillations. Before analyzing such forces, let's write the analogous equation for the longitudinal dynamics.

### 3.2    The synchrotron motion

In the longitudinal case, the motion is governed by the RF voltage, which we write as

$$V(t) = \hat{V} \sin\left[\omega_{rf} t + \varphi_s\right] , \tag{24}$$

where $\varphi_s$ is the synchronous phase and $\omega_{rf}$ the RF frequency. In the linear approximation of the RF voltage around the synchronous phase, and in the absence of the self-induced forces, the equation of motion is that of a simple harmonic oscillator

$$z'' + \left(\frac{Q_z}{\rho_x}\right)^2 z = 0 \tag{25}$$

with particles oscillating around the synchronous phase $\varphi_s$ with a synchronous tune given by:

$$Q_z = \frac{\omega_z}{\omega_0} = \sqrt{\frac{qh\eta \hat{V} \cos\varphi_s}{2\pi\beta^2 E_0}} \tag{26}$$

with $h$ the harmonic number, $E_0$ the machine nominal energy, and

$$\eta = \frac{1}{\gamma^2} - \alpha_c . \tag{27}$$

The slippage factor $\eta$ accounts for the increase of the speed with energy ($1/\gamma^2$), and the length of the real orbit due to the dispersion ($\alpha_c$).

If we also include in Eq. (25) the longitudinal self-induced forces due to the interaction of the charge with the surroundings, we obtain:

$$z'' + \left(\frac{Q_z}{\rho_x}\right)^2 z = \frac{\eta \, F_z^{self}(s)}{m_0 v_z^2 \gamma} . \tag{28}$$

Which represents the longitudinal equation of motion corresponding to the Eqs. (22) and (23).

## 4. Space charge forces

### 4.1 Direct space charge forces in the free space

Let us consider a relativistic charge moving with constant velocity $\vec{v}$. It is well known that its electrostatic field is modified because of the relativistic Lorentz contraction factor along the direction of motion as shown in Fig. 3. For an ultra-relativistic charge with $\gamma^2 \gg e$, the field lines are confined on a plane perpendicular to the direction of motion.

$$E_x(z=0) = \frac{q}{4\pi\varepsilon_0} \frac{\gamma x}{\left[x^2 + y^2\right]^{3/2}}$$

$$E_y(z=0) = \frac{q}{4\pi\varepsilon_0} \frac{\gamma y}{\left[x^2 + y^2\right]^{3/2}}$$

$$E_z(x=y=0) = \frac{q}{4\pi\varepsilon_0} \frac{\gamma z}{\left[\gamma^2 z^2\right]^{3/2}} = \frac{q}{4\pi\varepsilon_0} \frac{1}{\gamma^2 z^2}$$

$$\vec{E}_\rho = \frac{q}{4\pi\varepsilon_0} \frac{\gamma \vec{\rho}}{\rho^3}$$

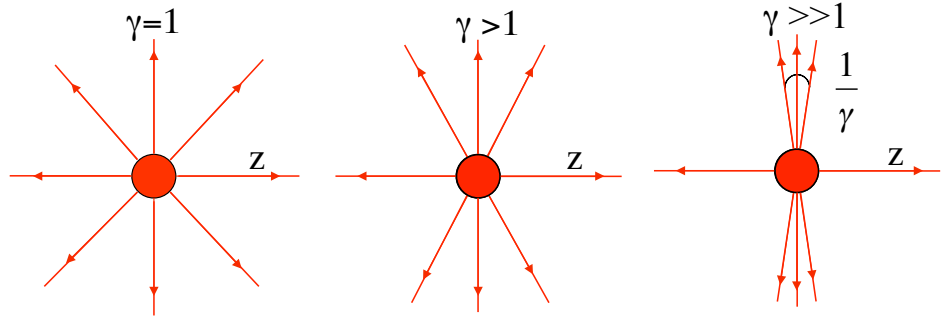

**Fig. 3:** Field lines for charges at different energies.

If now another charge is travelling on a parallel trajectory with respect to the first one along the z-axis, it is easy to see that the e.m. forces between them vanish. In fact, the electric and magnetic fields of a point charge moving with constant velocity $v = \beta c$, can be written as:

$$E_r(z=0) = \frac{q}{4\pi\varepsilon_0} \frac{\gamma}{r^2}$$

$$B_\phi(z=0) = \frac{q\beta}{4\pi\varepsilon_0 c} \frac{\gamma}{r^2} \qquad . \qquad (29)$$

$$E_z(r=0) = \frac{q}{4\pi\varepsilon_0} \frac{1}{\gamma^2 z^2}$$

If the two charges travel along the z axis at $r = 0$ with different longitudinal positions, then the force is proportional to the longitudinal electric field $E_z$, which vanishes as $1/\gamma^2$. On the other hand, if the charges have the same longitudinal position ($z = 0$) and different transverse positions, due to the combined effect of the defocusing electric and focusing magnetic fields we get

$$F_r = q\left(E_r - \beta c B_\phi\right) = \frac{q\gamma}{4\pi\varepsilon_0 r^2}\left(1-\beta^2\right) = \frac{q}{4\pi\varepsilon_0 \gamma} \frac{1}{r^2} \quad . \tag{30}$$

In both cases, for $\gamma \to \infty$, a charge travelling close to another one on a parallel trajectory is not affected by e.m. forces.

Let us now consider the case of a uniform cylindrical charge distribution travelling with ultra-relativistic speed in the free space. Under these assumptions, the electric field lines are perpendicular to the direction of motion, and the magnetic ones are circumferences, as shown in Fig. 4. The transverse electric and magnetic fields intensity can be computed as in the static case, applying the Gauss and Ampere laws:

$$\int_S \vec{E} \cdot \hat{n} \, dS = \frac{q}{\varepsilon_0}, \qquad \oint_l \vec{B} \cdot d\vec{l} = \mu_0 I \quad . \tag{31}$$

We now suppose that the beam is a uniform cylinder of radius *a* so that the longitudinal charge distribution (charge per unit of length) can be written as $\lambda(r) = \lambda_o \left(\frac{r}{a}\right)^2$, and we want to compute the transverse space charge forces acting on a particle inside the beam.

Applying Eqs. (31) to a cylinder for Gauss law and to a circumference for Ampere law, we obtain:

$$E_r(2\pi r)\Delta z = \frac{\lambda(r)\Delta z}{\varepsilon_0} \Rightarrow E_r = \frac{\lambda(r)}{2\pi\varepsilon_0 r} = \frac{\lambda_0}{2\pi\varepsilon_0}\frac{r}{a^2}$$
$$2\pi r B_\phi = \mu_0 J \pi r^2 = \mu_0 \beta c \lambda(r) \Rightarrow B_\phi = \frac{\lambda_o \beta}{2\pi\varepsilon_0 c}\frac{r}{a^2} \quad . \quad (32)$$

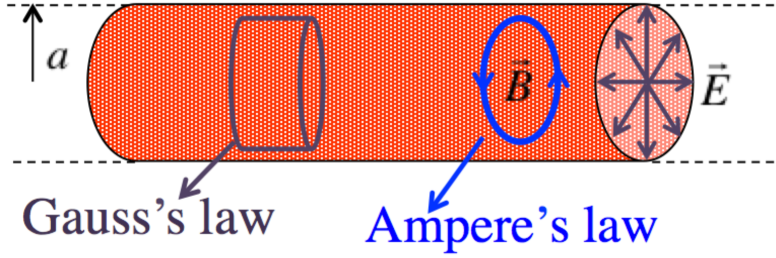

**Fig. 4:** Uniform cylindrical charge distribution with its electric and magnetic fields.

We now observe that $B_\phi = \frac{\beta}{c}E_r$, so that the e.m. transverse force acting on a charge inside the beam is

$$F_r(r) = e(E_r - \beta c B_\phi) = e(1-\beta^2)E_r = \frac{e}{\gamma^2}\frac{\lambda_o}{2\pi\varepsilon_0}\frac{r}{a^2} \quad . \quad (33)$$

We can therefore conclude that inside a uniform cylindrical charge density, travelling with ultra-relativistic speed, the transverse space charge forces vanish as $1/\gamma^2$ due to the cancellation of the electric and the magnetic forces.

### 4.2  Effects of conducting and magnetic screens

In an accelerator, the beams travel inside a vacuum pipe generally made of metallic material (such as aluminum, copper, or stainless steel). This pipe also passes through the coils of magnets (dipoles, quadrupoles, sextupoles), and its cross-section may have a complicated shape, as in the case of special devices like RF cavities, kickers, diagnostics and controls. However, most part of the beam pipe has a cross-section with a simple shape: circular, elliptic or racetrack. In order to obtain the space charge forces acting on a beam, let us consider here a smooth, perfectly conducting beam pipe.

Before dealing with the problem, it is first necessary to review the basic features of e.m. fields close to metallic and magnetic materials. A discussion of the boundary conditions is given in the Appendix 1. Here we report only the more relevant conclusions. The electric field of a point charge close to a conducting screen can be derived through the method of image charges, as shown in Fig. A1. For what concerns the magnetic field, a constant current close to a good conductor screen with $\mu_r \approx 1$, as copper or aluminium, produces circular field lines not affected by the presence of the material itself. However, if the material is of ferromagnetic type, with $\mu_r \gg 1$, due to its magnetisation, the magnetic field lines are strongly affected, inside and outside the material. In particular a very high magnetic permeability makes the tangential field zero at the boundary so that the total magnetic field must be perpendicular to the surface, just like the electric field lines close to a conductor (see Fig. A2).

The scenario changes when we deal with time-varying fields for which it is necessary to compare the wall thickness and the skin depth (region of penetration of the e.m. fields) in the conductor. If the fields penetrate and pass through the material, we are practically in the quasi-static boundary conditions case, depending on the (negligible) effect of the outermost material. Conversely, if the skin depth is very small, fields do not penetrate, and then the electric field lines are perpendicular to the wall, as in the static case, while the magnetic field lines are tangent to the surface. In this case, the magnetic field lines can be obtained by considering two currents flowing in opposite directions, as discussed in Appendix A1.2.

In the following paragraphs, we analyse the space charge forces that originates in presence of the screens in some simple cases.

### 4.3 Circular perfectly conducting pipe with the beam at center and direct space charge forces

Due to the symmetry, the transverse e.m. fields produced by an ultra-relativistic charge moving on the axis of a circular, perfectly conducting pipe are the same as in the free space. This implies that for a charge distribution with cylindrical symmetry, the total force acting on a charge inside the beam is still given by Eq. (33). It is interesting to note that, in the ultra-relativistic regime, this result does not depend on the longitudinal distribution of the beam, so that, considering more generally a uniform radial distribution and a longitudinal line density λ(z), the radial force can be written as

$$F_r(r,z) = \frac{e}{\gamma^2} \frac{\lambda(z)}{2\pi\varepsilon_0} \frac{r}{a^2}. \tag{34}$$

This force, as that in the free space, depends on $1/\gamma^2$ due to the cancellation of the electric and magnetic forces, and it is linear with the transverse position $r$. This result depends on the hypothesis of uniform transverse distribution. Indeed, if the transverse distribution is not constant, we can still apply Gauss's law to obtain the electric field and Ampere's law to obtain the magnetic field. For example, given the following bi-Gaussian distribution:

$$\rho(r,z) = \frac{q_0}{\left(\sqrt{2\pi}\right)^3 \sigma_z \sigma_r^2} e^{\frac{-z^2}{2\sigma_z^2}} e^{\frac{-r^2}{2\sigma_r^2}} \tag{35}$$

with $q_0$ the bunch charge, the Gauss's law applied to a cylinder as that of Fig. 4 with an infinitesimal height $dz$, gives, as a radial electric field (we suppose $\gamma \to \infty$ so that $E_z \approx 0$),

$$E_r(r,z) = \frac{1}{2\pi\varepsilon_0} \frac{q_0}{\sqrt{2\pi}\sigma_z \sigma_r^2 r} e^{\frac{-z^2}{2\sigma_z^2}} \int_0^r e^{\frac{-r'^2}{2\sigma_r^2}} r' dr' = \frac{1}{2\pi\varepsilon_0} \frac{q_0}{\sqrt{2\pi}\sigma_z} e^{\frac{-z^2}{2\sigma_z^2}} \left[\frac{1-e^{\frac{-r^2}{2\sigma_r^2}}}{r}\right]. \tag{36}$$

The magnetic field can be obtained in the same way as in Eq. (32) so that the total force on a charge inside the bunch is

$$F_r(r,z) = e\left(1-\beta^2\right)E_r = \frac{e}{2\pi\varepsilon_0\gamma^2} \frac{q_0}{\sqrt{2\pi}\sigma_z} e^{\frac{-z^2}{2\sigma_z^2}} \left[\frac{1-e^{\frac{-r^2}{2\sigma_r^2}}}{r}\right]. \tag{37}$$

It is also important to observe that the self-induced forces given by Eqs. (34) and (37) are always defocusing both in *x* and *y* directions, as shown in Fig. 5. Note also that the force given by Eq. (37) is not linear in the transverse position $r$.

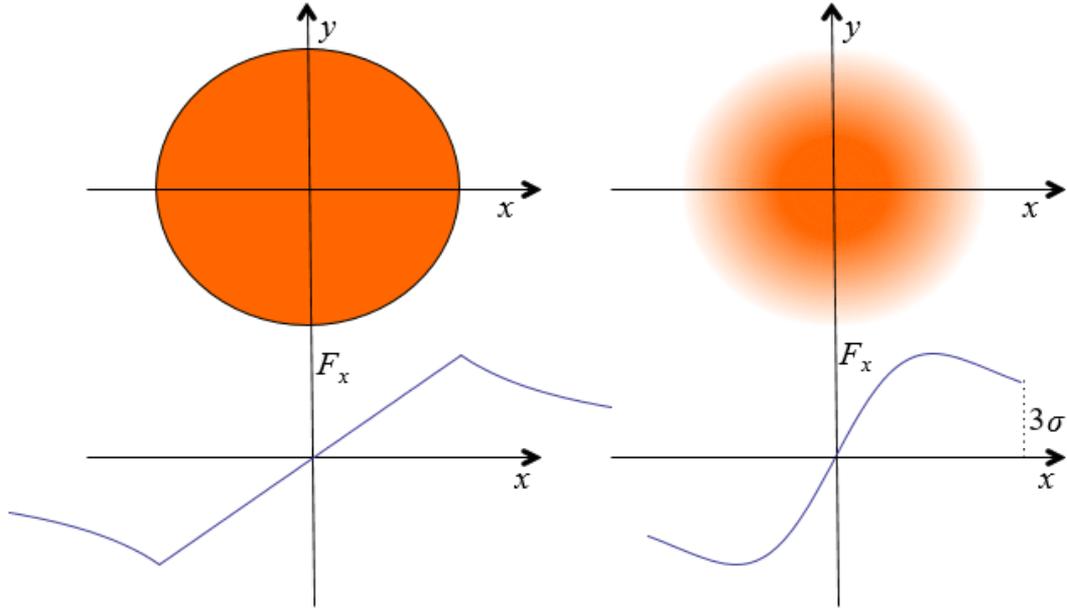

**Fig. 5:** Defocusing transverse self-induced forces produced by direct space charge in case of uniform (left) and Gaussian (right) distributions.

### 4.4 Parallel plates with the beam at center and indirect space charge forces for d.c. currents

In some cases, also with elliptical and rectangular beam pipes, the cross-section is such that we can consider only the surfaces closer to the beam, which can be approximated by two parallel plates. Let us suppose to have a cylindrical charge distribution of radius *a* and line density *λ(z)* between two conducting plates 2*h* apart. For obtaining the static electric field, the two conducting plates can be removed by using the method of image charges and substituted by an infinite series of charges with alternating signs 2*h* apart from each other, as shown in Fig. 6.

We want to evaluate the electric field, due to the image charges, at a position *y* inside the bunch (*y<a*). The transverse field produced by the image charge distribution immediately above the real one can be written as

$$E_y^{1,up,im}(z,y) = \frac{\lambda(z)}{2\pi\,\varepsilon_0} \frac{1}{2h-y} \tag{38}$$

while the transverse electric field of the image charge distribution immediately below the real one is

$$E_y^{1,down,im}(z,y) = -\frac{\lambda(z)}{2\pi\,\varepsilon_0} \frac{1}{2h+y} \; . \tag{39}$$

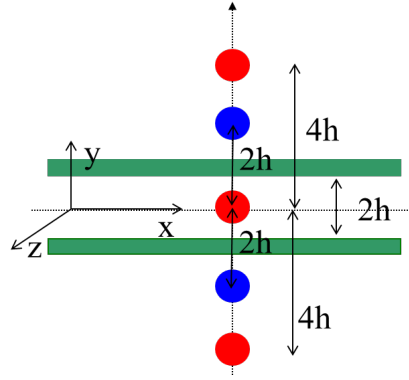

**Fig. 6:** A charge between two parallel plates and its image charges.

If we sum the contribution of all the infinite image charge distributions, we get the total transverse electric field

$$E_y^{im}(z,y) = \frac{\lambda(z)}{2\pi\,\varepsilon_0}\sum_{n=1}^{\infty}(-1)^n\left[\frac{1}{2nh+y}-\frac{1}{2nh-y}\right] = \frac{\lambda(z)}{2\pi\,\varepsilon_0}2y\sum_{n=1}^{\infty}(-1)^{n+1}\frac{1}{(2nh)^2-y^2}. \quad (40)$$

The sum on the right-hand side of the equation can be done under the simplified hypothesis that $h>>a>y$. Under this condition, we can ignore the term $y$ in the denominator of the sum and we get

$$E_y^{im}(z,y) \cong \frac{\lambda(z)}{2\pi\,\varepsilon_0}\frac{y}{2h^2}\sum_{n=1}^{\infty}\frac{(-1)^{n+1}}{n^2} = \frac{\lambda(z)}{4\pi\,\varepsilon_0 h^2}\frac{\pi^2}{12}y. \quad (41)$$

For d.c. or slowly varying currents, we have seen that the boundary conditions imposed by the conducting walls do not affect the magnetic field, which remains circular with no image currents. As a consequence, there is no cancellation effect of the electric and magnetic forces for the fields produced by the image charges, differently from what we have obtained for the real charges (direct forces). Therefore, the indirect space charge force acting on a charge inside the beam is generated only by the electric field (41) times the particle charge.

From the divergence equation, we can also derive the other transverse component of the electric field along $x$

$$\frac{\partial}{\partial x}E_x^{im} = -\frac{\partial}{\partial y}E_y^{im} \Rightarrow E_x^{im}(z,x) = \frac{-\lambda(z)}{4\pi\,\varepsilon_0 h^2}\frac{\pi^2}{12}x. \quad (42)$$

From the above fields, the total forces acting on a charge inside the bunch moving between two parallel plates, including also the direct space charge force given by Eq. (34), are:

$$F_x(z,x) = \frac{e\lambda(z)}{\pi\,\varepsilon_0}\left(\frac{1}{2a^2\gamma^2}-\frac{\pi^2}{48h^2}\right)x \quad (43)$$

$$F_y(z,y) = \frac{e\lambda(z)}{\pi\,\varepsilon_0}\left(\frac{1}{2a^2\gamma^2}+\frac{\pi^2}{48h^2}\right)y. \quad (44)$$

Therefore, for $\gamma>>1$, and for d.c. or slowly varying currents, the cancellation effect applies only to the direct space charge forces. There is no cancellation of the electric and magnetic forces due to the image charges.

## 4.5 Parallel plates with the beam at center and indirect space charge forces for a.c. currents

We have seen that close to a conductor the e.m. fields have different behaviours, depending on the skin depth $\delta_w$ of the material as discussed in Appendix 1. Usually, the frequency spectrum of a beam is quite rich in harmonics, especially for bunched beams. It is then convenient to decompose the current into a d.c. component, $\bar{I}$, for which $\delta_w \gg \Delta w$, with $\Delta w$ the width of the beam pipe, and an a.c. component, $\hat{I}$, for which $\delta_w \ll \Delta w$. While the d.c. component of the magnetic field does not perceive the presence of the material, so we can apply Eqs. (43) and (44), its a.c. component produces a magnetic field tangent to the wall, which can be obtained by using an infinite sum of image currents with alternating directions. In this case, to get the total magnetic field of the image currents, we can follow the same procedure we have used for the electric fields given by Eqs. (41) and (42), and by considering the relation between a.c. current and its charge distribution: $\hat{I} = \beta c \hat{\lambda}$, thus obtaining a magnetic field due to image currents of the kind

$$\hat{B}_x(z,y) = -\frac{\beta}{c}\hat{E}_y(z,y) = -\frac{\beta^2 \hat{\lambda}(z)}{\pi \varepsilon_0}\frac{\pi^2}{48 h^2}y \ . \tag{45}$$

We can see from the above expression that, in this case, the attractive magnetic force tends to compensate for the repulsive electric one, which is still given by the second term in the parentheses of Eq. (44), so that we have a total force due to image charges and currents given by

$$\hat{F}_y(z,y) = \frac{e\hat{\lambda}(z)}{\pi \ \varepsilon_o \gamma^2}\frac{\pi^2}{48 h^2}y \ . \tag{46}$$

Combining Eq. (46) with the direct space charge force, we get for the a.c. component:

$$\hat{F}_y(z,y) = \frac{e\hat{\lambda}(z)}{2\pi \ \varepsilon_0 \gamma^2}\left(\frac{1}{a^2}+\frac{\pi^2}{24 h^2}\right)y \ . \tag{47}$$

Analogously, along the *x* direction, we get

$$\hat{F}_x(z,x) = \frac{e\hat{\lambda}(z)}{2\pi \ \varepsilon_0 \gamma^2}\left(\frac{1}{a^2}-\frac{\pi^2}{24 h^2}\right)x \ . \tag{48}$$

## 4.6 Parallel plates with the beam at center and indirect space charge forces for d.c. currents in the presence of ferromagnetic materials

As a last example, we consider now the case where, outside of the metallic pipe, there is a dipole magnet. The magnetic field produced by the d.c. currents $\beta c \bar{\lambda}(z)$ doesn't see the conducting pipe, while it is strongly affected by ferromagnetic material. In fact, as seen in Appendix 1, the magnetic field lines must be orthogonal to the pole surface. We have also seen that the total magnetic field can be obtained by removing the screen and considering image currents flowing in the same direction.

Proceeding analogously to Section 4.4, by denoting with *g* the gap in the dipole magnet, we obtain:

$$B_x^{im}(z,y) = \frac{\mu_0 \beta c \bar{\lambda}(z)}{2\pi}\sum_{n=1}^{\infty}\left[\frac{1}{2ng-y}-\frac{1}{2ng+y}\right] \ . \tag{49}$$

Note that, in this case, we don't have the term $(-1)^n$ in the sum. By using the same approximation as before: $h \gg a > y$, we obtain a magnetic field due to the image currents equal to

$$B_x^{im}(z,y) \cong \frac{\mu_0 \beta c \bar{\lambda}(z) y}{4\pi g^2} \sum_{n=1}^{\infty} \frac{1}{n^2} = \frac{\mu_0 \beta c \bar{\lambda}(z) \pi^2 y}{24\pi g^2} \tag{50}$$

and the corresponding force is

$$F_y^{im}(z,y) = \frac{\beta^2 \bar{\lambda}(z) \pi^2}{24\pi \varepsilon_0 g^2} y . \tag{51}$$

In order to obtain the magnetic field acting on a particle displaced along the *x* direction, we can use the relation $\vec{\nabla} \times \vec{B} = 0$, which, for the *z* direction, gives

$$\frac{\partial B_y}{\partial x} = \frac{\partial B_x}{\partial y} \tag{52}$$

so that

$$B_y^{im}(z,x) = \frac{\mu_0 \beta c \bar{\lambda}(z) \pi^2}{24\pi g^2} x \tag{53}$$

and

$$F_x^{im}(z,x) = -\frac{\beta^2 \bar{\lambda}(z) \pi^2}{24\pi \varepsilon_0 g^2} x . \tag{54}$$

### 4.7 Parallel plates with the beam at center: general expression of the force

Taking into account all the boundary conditions and either d.c. and a.c. currents, we can summarize what we have obtained in the previous sections for the parallel plates, and write the general expression of the force as:

$$F_u = \frac{e}{2\pi \ \varepsilon_0} \left[ \frac{1}{\gamma^2} \left( \frac{1}{a^2} \mp \frac{\pi^2}{24h^2} \right) \lambda \mp \beta^2 \left( \frac{\pi^2}{24h^2} + \frac{\pi^2}{12g^2} \right) \bar{\lambda} \right] u \tag{55}$$

where $\lambda$ is the total current divided by $\beta c$, and $\bar{\lambda}(z)$ its d.c. term. We take the sign (+) if *u=y*, and the sign (–) if *u=x*.

One often finds the above expression written as:

$$F_u = \frac{e}{\pi \ \varepsilon_0} \left[ \frac{1}{\gamma^2} \left( \frac{\varepsilon_0}{a^2} \mp \frac{\varepsilon_1}{h^2} \right) \lambda \mp \beta^2 \left( \frac{\varepsilon_1}{h^2} + \frac{\varepsilon_2}{g^2} \right) \bar{\lambda} \right] u \tag{56}$$

where the Laslett form factors [10] $\varepsilon_0$, $\varepsilon_1$ and $\varepsilon_2$ can be obtained for some different beam pipe geometries. For example, for parallel plates, by comparing Eq. (55) with (56) we get $\varepsilon_0 = 1/2$, $\varepsilon_1 = \pi^2/48$, $\varepsilon_2 = \pi^2/24$ [11]. It is interesting to note that these forces are anyway linear in the transverse displacement *x* and *y* due to the assumption of uniform transverse distribution.

### 4.8 Longitudinal direct space charge force

Up to now, we have obtained the transverse forces, direct and indirect, produced by space charge distributions. The longitudinal electric field, responsible for the longitudinal forces, can be derived starting from the knowledge of the transverse fields, as shown in Appendix 2. The transverse electric field inside the beam ($r \leq a$) can be expressed by the first of Eqs. (32) for a uniform transverse distribution, which, however, can be generalized by considering a non-uniform longitudinal distribution $\lambda(z)$. This field, outside the beam, for $r \geq a$ is equal to

$$E_r(r \geq a) = \frac{\lambda(z)}{2\pi\varepsilon_0 r}. \tag{57}$$

As a consequence, the last equation of the Appendix 2 becomes

$$E_z(r,z) = -\frac{1}{2\pi\varepsilon_0 \gamma^2}\left[\int_r^a \frac{r'}{a^2}dr' + \int_a^b \frac{1}{r'}dr'\right]\frac{\partial \lambda(z)}{\partial z} \tag{58}$$

giving a longitudinal force of the kind

$$F_z(r,z) = \frac{-e}{4\pi\varepsilon_0 \gamma^2}\left(1 - \frac{r^2}{a^2} + 2\ln\frac{b}{a}\right)\frac{\partial \lambda(z)}{\partial z}. \tag{59}$$

Therefore, the longitudinal force acting on a charge is positive (negative) in the region with a negative (positive) density slope.

## 5. Coherent and incoherent tune shifts

### 5.1 Coherent and incoherent effects

We are now ready to study the effects of the space charge forces on the beam dynamics. When the beam is located at the centre of symmetry of the pipe, the e.m. forces due to direct and image space charge cannot affect the motion of the centre of mass (called coherent motion), but they change the trajectory of individual charges inside the beam (incoherent effects). These forces may have a complicated dependence on the charge position. A simple analysis is done considering only the linear expansion of the self-induced forces around the equilibrium trajectory. Observe that the forces given by Eq. (34) or (55) are already linear in the transverse coordinate $r$, and, in such cases, there's no need for linear expansion.

Referring to the equations of motion (22), (23), and (28), let's then expand the self-induced forces around the ideal orbit analogously to what we have done for the external forces. A constant term in the expansion of $F^{self}$ changes the equilibrium orbit in the transverse plane and the synchronous phase in the longitudinal one, while the linear term, proportional to the charge displacement, changes the focusing strength and, therefore, induces a shift of the betatron and synchrotron frequencies.

This can happen both in the motion of individual particles inside the beam (incoherent motion) and in the transverse oscillations of the whole beam (coherent motion) around the closed orbit when the beam is off-center with respect to the beam pipe.

### 5.2 Transverse incoherent effects

Let us consider only the linear term in the transverse self-induced forces, that is

$$F_x^{self}(x,s) \cong \left(\frac{\partial F_x^{self}}{\partial x}\right)_{x=0} x. \tag{60}$$

In the case of uniform transverse beam distribution, both in a circular pipe and between parallel plates, the force given by Eq. (60) is not an approximation, as shown by Eqs. (34) and (55). For other kinds of distributions, such as the one given by Eq. (35), we can always suppose that the transverse displacement $r$ of a charge is much smaller than the transverse bunch dimension $\sigma_r$, so that the term inside the square brackets in Eq. (37) can be expanded at first order in $r$ giving $(r/2\sigma)$. In any case, we end up with a self-induced force linearly dependent on the particle displacement. As a consequence, by considering, for example, the motion along the $x$ direction, Eq. (22) becomes

$$x''(s) + \left(\frac{Q_x}{\rho_x}\right)^2 x(s) = \frac{1}{m_0 v_z^2 \gamma}\left(\frac{\partial F_x^{self}}{\partial x}\right)_{x=0} x(s). \tag{61}$$

The linear additional term on the right-hand side of the equation produces a shift of the betatron tune $Q_x$. Indeed, we can write

$$x''(s) + \left[\left(\frac{Q_x}{\rho_x}\right)^2 - \frac{1}{\beta^2 E_0}\left(\frac{\partial F_x^{self}}{\partial x}\right)_{x=0}\right] x(s) = 0, \tag{62}$$

where the term $m_0 v_z^2 \gamma$ has been substituted with $\beta^2 E_0$ by approximating $v_z \simeq v$. We now recognize in the brackets a term proportional to the square of the new betatron tune, shifted, with respect to the initial one $Q_x$, by the self-induced forces. This new tune can be written as $(Q_x + \Delta Q_x)^2 / \rho_x^2$. For small perturbations, the betatron tune shift $\Delta Q_x$ can then be obtained by:

$$\left[\left(\frac{Q_x}{\rho_x}\right)^2 - \frac{1}{\beta^2 E_0}\left(\frac{\partial F_x^{self}}{\partial x}\right)_{x=0}\right] = \frac{(Q_x + \Delta Q_x)^2}{\rho_x^2} \cong \frac{Q_x^2 + 2Q_x \Delta Q_x}{\rho_x^2} \tag{63}$$

thus giving

$$\Delta Q_x = -\frac{\rho_x^2}{2\beta^2 E_0 Q_x}\left(\frac{\partial F_x^{self}}{\partial x}\right). \tag{64}$$

A similar expression is found in the $y$ direction. The betatron tune shift is negative since the space charge forces are always defocusing on both planes. Note that the tune shift is, in general, a function of $z$, due to the dependence of the self-induced force on $\lambda(z)$. The consequence is a tune spread inside the beam. This conclusion is generally true also for more realistic non-uniform transverse beam distributions, which are characterized by a tune shift dependent also on the betatron oscillation amplitude. In these cases, instead of tune shift, the effect is called tune spread.

As an example of the application of the above expression, let's find the incoherent shift of the betatron tune for a uniform electron beam of charge $eN_p$, radius $a$ and length $l_0$, inside a circular pipe. The self-induced force is given by Eq. (34), with $\lambda(z) = eN_p / l_0$ so that

$$\Delta Q_x = -\frac{\rho_x^2 e^2 N_p}{4\pi\varepsilon_0 a^2 l_0 \beta^2 \gamma^2 E_0 Q_x} \tag{65}$$

or, expressed in terms of the classical radius of the electron $r_0$:

$$\Delta Q_x = -\frac{r_0 \rho_x^2 N_p}{a^2 l_0 \beta^2 \gamma^3 Q_x}. \tag{66}$$

In the general case of non-uniform focusing along the accelerator, as given by Eq. (16), the linear effect of the self-induced forces can be treated as a quadrupole error $\Delta K_u$[11] distributed along the accelerator, with $u$ representing one of the axes $x$ or $y$, thus giving a betatron tune shift of:

$$\Delta Q_u = \frac{1}{4\pi}\oint \beta_u(s)\Delta K_u(s)ds = \frac{-1}{4\pi\beta^2 E_0}\oint \beta_u(s)\left(\frac{\partial F_u^{self}}{\partial u}\right)ds. \qquad (67)$$

For example, by considering the previous case of a uniform electron beam inside a circular pipe, but with non-uniform focusing, Eq. (65) will be replaced by

$$\Delta Q_x = -\frac{r_0 N_p}{2\pi\beta^2\gamma^3 l_0}\oint \frac{\beta_x(s)}{a^2(s)}ds = -\frac{r_0 N_p}{2\pi\beta^2\gamma^3 l_0}\frac{2\pi\rho_x}{\varepsilon_x}, \qquad (68)$$

which has been obtained by observing that the quantity $a^2(s)/\beta_x(s)$ is the beam emittance $\varepsilon_x$, which is constant along the machine.

### 5.3  Transverse coherent effects

Suppose we have a beam that, because, for example, of coherent betatron oscillations, is displaced from the pipe axis. Due to induction, there will be a higher concentration of charges of opposite sign on the pipe surface closer to the beam. These induced charges attract the beam more intensely than the induced charges on the opposite side. As a consequence, the beam center of mass will experience a defocusing force.

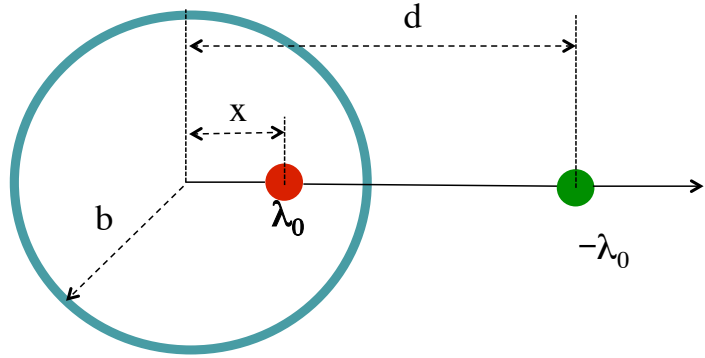

**Fig. 7:** A charge distribution off-axis inside a cylindrical pipe and its image charge distribution.

As example, let us suppose to have an electron beam with uniform longitudinal charge distribution $\lambda_0 = eN_p/l_0$ inside a conducting cylindrical pipe of radius $b$, displaced by $x$ from its axis, as shown in Fig. 7. Since the electric field lines of the charge distribution must be perpendicular to the pipe, by imposing that the conductor surface is equipotential, we can remove the walls and substitute the pipe with a charge distribution $-\lambda_0$ at a distance from the axis equal to $d = b^2/x$ on the same side of the displacement.

The image charge distribution attracts the whole beam thus producing a coherent defocusing effect. The electric field of the image charge acting on the center of mass of the beam is

$$E_x^{im}(x) \cong \frac{\lambda_0}{2\pi\,\varepsilon_0}\frac{1}{d-x}. \qquad (69)$$

If we linearize the electric field by considering small displacements of the beam, such that $x<<d$, then $1/(d-x) \cong 1/d = x/b^2$, and we obtain

$$E_x^{im}(x) \cong \frac{\lambda_0}{2\pi \, \varepsilon_0} \frac{x}{b^2}, \tag{70}$$

which gives a linear coherent force

$$F_x^{self}(x) \cong \frac{e\lambda_0}{2\pi \, \varepsilon_0} \frac{x}{b^2}. \tag{71}$$

This force produces a coherent betatron tune shift that can still be evaluated by using Eq. (64) giving

$$\Delta Q_x = -\frac{\rho_x^2 e \lambda_0}{4\pi \, \varepsilon_0 \beta^2 E_0 b^2 Q_x} = -\frac{r_0 \rho_x^2 N_p}{b^2 l_0 \beta^2 \gamma Q_x} \ . \tag{72}$$

This coherent betatron tune shift is different from the incoherent one given by Eq. (66) since it does not depend on the beam size but on the pipe radius, and it is inversely proportional to the beam energy.

**5.4   Longitudinal incoherent effects**

The effects of longitudinal space charge forces on the beam dynamics can be obtained by using Eq. (28). We expand the self-induced force and, as already discussed in Section 5.1, we observe that the constant term in the expansion leads to a shift of the synchronous phase, while the linear term, proportional to the displacement, changes the focusing strength and therefore induces a shift of the synchrotron tune.

Let us analyse the motion of the beam around the new equilibrium phase and let us consider the linear term of the longitudinal self-induced force:

$$F_z^{self}(r,z) \cong \left(\frac{\partial F_z^{self}}{\partial z}\right)_{z=0} z \, . \tag{73}$$

The equation of motion (28) becomes

$$z'' + \left(\frac{Q_z}{\rho_x}\right)^2 z = \frac{\eta}{\beta^2 E_0} \left(\frac{\partial F_z^{self}}{\partial z}\right)_{z=0} z \, . \tag{74}$$

As in the transverse case, let us consider the same approximation leading to Eq. (63), obtaining in this case

$$\Delta Q_z = \frac{-\eta \rho_x^2}{2\beta^2 E_0 Q_z} \left(\frac{\partial F_z^{self}}{\partial z}\right) \, . \tag{75}$$

Differently from the transverse betatron tune shifts, the synchrotron tune shift can be either positive or negative and changes with the position of the charge inside the beam.

For example, let us consider a transverse uniform beam of radius $a$ in a cylindrical pipe, having a parabolic longitudinal distribution of the kind

$$\lambda(z) = \frac{3eN_p}{2l_0}\left[1-\left(\frac{2z}{l_0}\right)^2\right]. \tag{76}$$

The incoherent tune shift can be obtained by combining Eqs. (59), (75) and (76), thus giving

$$\Delta Q_z = -6\frac{\eta \rho_x^2 r_0 N_p}{\beta^2 \gamma^3 Q_z}\left(1 - \frac{r^2}{a^2} + 2\ln\frac{b}{a}\right)\left(\frac{1}{l_0^3}\right), \tag{77}$$

which depends on the transverse position $r$ of the charge inside the beam. If, instead of a parabolic bunch distribution, we had a Gaussian one, then the synchrotron tune shift would have a dependence also on the longitudinal position $z$.

## 6. Consequences of the space charge tune shifts

In circular accelerators the values of the betatron tunes should not be close to rational numbers in order to avoid the crossing of linear and non-linear resonances where the beam becomes unstable. The tune spread induced by the space charge force can make hard to satisfy this basic requirement. Typically, in order to avoid major resonances the stability requires [9, 12]

$$|\Delta Q_u| < 0.5.$$

If the tune spread exceeds this limit, it is possible to reduce the effect of space charge tune spread by increasing the injection energy.

It is worth noting that the incoherent tune spread produces also a beneficial effect, called Landau damping [13], which can cure the coherent instabilities, provided that the coherent tune remains inside the incoherent spread.

# Appendix 1 – Boundary Conditions for Conductors

## A1.1 Static Electric and Magnetic Fields

When we have two materials with different relative permittivity, which we call $\varepsilon_{r1}$ and $\varepsilon_{r2}$, and there is no surface charge distribution at the boundary between the two materials, in the passage from one to the other, the tangential electric field and the normal electric displacement are preserved so that we have the boundary relations:

$$E_{t1} = E_{t2}$$
$$\varepsilon_{r1} E_{n1} = \varepsilon_{r2} E_{n2}.$$

If one of the two materials is a conductor with a finite conductivity, then the electric field vanishes inside it, and the walls are equipotential surfaces. This implies that the electric field lines are orthogonal to the conductor surface, independently of the dielectric and magnetic properties of the material. The only condition is to have a finite conductivity.

If we have a charge close to a conductor, in order to obtain the electric field, we need to include the effects of the induced charges on the conducting surfaces, and we must know how they are distributed. Generally, this task is not easy, but if we have an infinite conducting plane screen, the problem can be easily solved by making use of the method of image charges: we can remove the screen and put at a symmetric location a charge with opposite sign, as shown in Fig. A1.

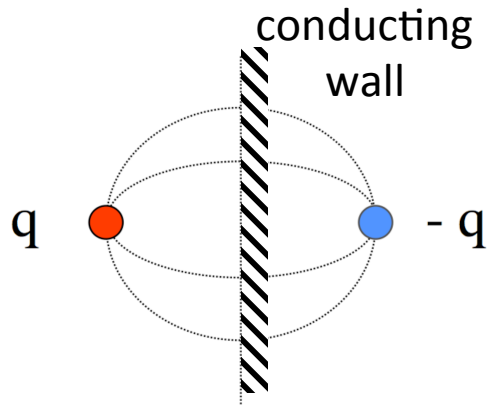

**Fig. A1:** Method of images.

The total electric field on the left side of the screen is the sum of the direct and the image fields:

$$\vec{E}^{tot} = \vec{E}^{direct} + \vec{E}^{images}.$$

For the static magnetic field between two materials with different permeability, the following boundary relations hold:

$$H_{t1} = H_{t2}$$
$$\mu_{r1} H_{n1} = \mu_{r2} H_{n2}.$$

Thus, static magnetic fields do not perceive the presence of the conductor, if it has a magnetic permeability $\mu_r \simeq 1$, as copper or aluminium, and the field lines expand as in the free space. However, a beam pipe in a real machine goes through many magnetic components (like dipoles and quadrupoles) made of ferromagnetic materials with high permeability (of the order of $10^3 - 10^5$). For these materials, due to the boundary conditions, the magnetic field lines are practically orthogonal to the surface. Similarly to electric field lines for a plane conductor, the total magnetic field can be derived by using the image method: we remove the ferromagnetic wall and put a symmetric current with the same sign, as shown in Fig. A2.

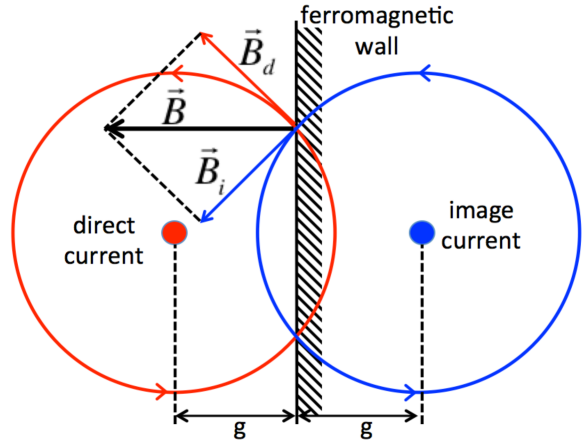

**Fig. A2:** Method of the image current.

## A1.2 Time-varying Fields

Static electric fields vanish inside a conductor for any finite conductivity, while static magnetic fields pass through unless of high permeability. This is no longer true for time-changing fields, which can penetrate inside the material in a region $\delta_w$ called skin depth. In order to obtain the skin depth as a function of the material properties, we write the following Maxwell's equations inside the conducting material together with the constitutive relations:

$$\begin{cases} \nabla \times \vec{E} = -\dfrac{\partial \vec{B}}{\partial t} \\ \nabla \times \vec{H} = \vec{J} + \dfrac{\partial \vec{D}}{\partial t} \end{cases} \qquad \begin{cases} \vec{B} = \mu \vec{H} \\ \vec{D} = \varepsilon \vec{E} \\ \vec{J} = \sigma \vec{E} \end{cases}.$$

Let us consider a plane wave linearly polarized with the electric field in the *y* direction propagating in the material along *x*, as shown in Fig. A3.

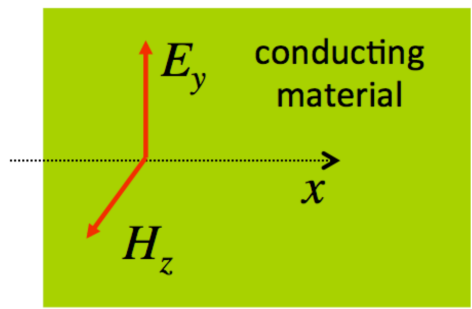

**Fig. A3:** Plane wave propagating inside a conducting material.

From Maxwell's equations, we get the wave equation for the electric field:

$$\frac{\partial^2 E_y}{\partial x^2} - \varepsilon\mu \frac{\partial^2 E_y}{\partial t^2} - \sigma\mu \frac{\partial E_y}{\partial t} = 0.$$

In order to find the solution of this wave equation, we assume that the electric field propagates in the *x* direction with the law

$$E_y = \tilde{E}_0 e^{i\omega t - \gamma x}.$$

If we substitute the above expression in the wave equation, we get the equation for the complex amplitude of the electric field $\tilde{E}_0$

$$(\gamma^2 + \varepsilon\mu\omega^2 - i\omega\mu\sigma)\tilde{E}_0 e^{i\omega t - \gamma x} = 0.$$

An analogous equation holds for $H_z$. In order to have non-zero electric field, the term inside the parentheses must be zero. If σ>>ωε this reduces to

$$\gamma \cong (1+i)\sqrt{\frac{\sigma\mu\omega}{2}}.$$

Under such a condition we say that the material behaves like a conductor. Since *γ* has a real part, fields propagating in the material are attenuated. The attenuation constant, measured in meters, is called skin depth $\delta_w$:

$$\delta_w \cong \frac{1}{\Re(\gamma)} = \sqrt{\frac{2}{\omega\sigma\mu}}.$$

The skin depth depends on the material properties and the frequency. Copper, for example, has a skin depth of

$$\delta_w \cong \frac{6.66}{\sqrt{f}}(cm).$$

If we assume a beam pipe 2 mm thick, we find that fields pass through the wall up to frequencies of about 1 kHz.

Time-varying fields generally pass through the conductor wall if the skin depth is larger than the wall thickness. This happens at relatively low frequencies when $\delta_w$ is large, while at higher frequencies, for a good conductor, the skin depth is very small. It can become much lower than the wall thickness, so we can consider that both electric and magnetic fields vanish inside the wall. In this condition, the electric field lines are perpendicular to the wall surface, as in the static case, while the magnetic field lines are tangent to the wall. As a consequence, in order to obtain the time-varying electric field of a beam close to a good conductor, we can still use the method of the image charges, while for the magnetic field, it is easy to see that we can use the method shown in Fig. A2, by changing the direction of the image current.

## Appendix 2 – Longitudinal Forces

In order to derive the relationship between the longitudinal and transverse forces inside a beam, let us consider the case of an ultra-relativistic bunch with cylindrical symmetry of radius $a$. We know from Faraday's law of induction that a varying magnetic field produces a rotational electric field:

$$\oint \vec{E} \cdot d\vec{l} = -\frac{\partial}{\partial t} \int_S \vec{B} \cdot \hat{n}\, dS .$$

In order to obtain the longitudinal electric field, we choose, as a path for the circulation, a rectangle going from a point inside the beam (at a distance $r$ from the axis) up to the beam pipe (a cylinder of radius $b$) with two sides parallel to the $z$-axis, as shown in Fig. A4. For a generic position $r < a$, and by taking $\Delta z$ small enough so that we can consider the electric field constant, we have:

$$E_z(r,z)\Delta z + \int_r^b E_r(r, z+\Delta z)dr - E_z(b,z)\Delta z - \int_r^b E_r(r,z)dr = -\Delta z \frac{\partial}{\partial t}\int_r^b B_\phi(r)dr .$$

We now write $E_r(r, z+\Delta z) - E_r(r,z) = \frac{\partial E_r(r,z)}{\partial z}\Delta z$ so that from the above equation we get

$$E_z(r,z) = E_z(b,z) - \int_r^b \left[\frac{\partial E_r(r,z)}{\partial z} + \frac{\partial B_\phi(r,z)}{\partial t}\right]dr .$$

By considering that $z = -\beta c t$, we can also write

$$E_z(r,z) = E_z(b,z) - \frac{\partial}{\partial z}\int_r^b \left[E_r(r,z) - \beta c B_\phi(r,z)\right]dr .$$

Since the transverse electric field and the azimuthal magnetic field are related by $B_\phi = \frac{\beta}{c}E_r$, we finally obtain

$$E_z(r,z) = E_z(b,z) - (1-\beta^2)\frac{\partial}{\partial z}\int_r^b E_r(r,z)dr .$$

Note that for perfectly conducting walls we have $E_z(b,z) = 0$, so that

$$E_z(r,z) = -\frac{1}{\gamma^2}\frac{\partial}{\partial z}\int_r^b E_r(r,z)dr .$$

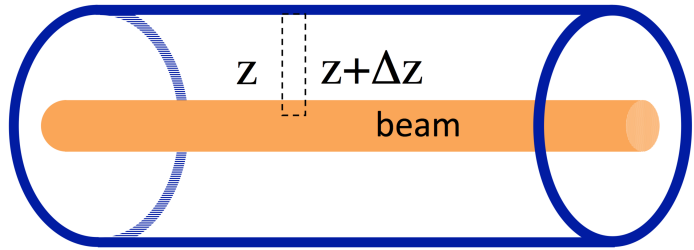

**Fig. A4:** Geometry for obtaining the longitudinal electric field due to space charge.